\documentclass[prl,aps,preprint,amsmath,superscriptaddress]{revtex4}
\usepackage{graphicx}
\usepackage{dcolumn}
\usepackage{bm}
\usepackage{epstopdf}
\usepackage{url}

\begin{document}
	
	\title{Metafluidic Tweezers: Precise Liquid Manipulation in Flowing Fluids}
	
	\author{Yuhong Zhou}
	\affiliation{Department of Physics, State Key Laboratory of Surface Physics, and Key Laboratory of Micro and Nano Photonic Structures (MOE), Fudan University, Shanghai 200438, China}
	
	\author{Fubao Yang}
	\affiliation{Graduate School of China Academy of Engineering Physics, Beijing 100193, China}
	
	\author{Jinrong Liu}
	\affiliation{Department of Physics, State Key Laboratory of Surface Physics, and Key Laboratory of Micro and Nano Photonic Structures (MOE), Fudan University, Shanghai 200438, China}
	
	\author{Gaole Dai}
	\affiliation{School of Physics and Technology, Nantong University, Nantong 226019, China}
	
	\author{Zixin Li}
	\affiliation{Department of Physics, State Key Laboratory of Surface Physics, and Key Laboratory of Micro and Nano Photonic Structures (MOE), Fudan University, Shanghai 200438, China}
	
	\author{Xuzhi Zhou}
	\affiliation{School of Aeronautics, Northwestern Polytechnical University, Xi'an, shaannxi 710072, China}
	
	\author{Peng Jin}\email{19110190022@fudan.edu.cn}
	\affiliation{Department of Physics, State Key Laboratory of Surface Physics, and Key Laboratory of Micro and Nano Photonic Structures (MOE), Fudan University, Shanghai 200438, China}
	
	\author{Jiping Huang}\email{jphuang@fudan.edu.cn}
	\affiliation{Department of Physics, State Key Laboratory of Surface Physics, and Key Laboratory of Micro and Nano Photonic Structures (MOE), Fudan University, Shanghai 200438, China}

\date{\today}
\begin{abstract}{\normalsize {\normalsize }}
Manipulating particles, such as cells and tissues, in a flowing liquid environment is crucial for life science research. Traditional contactless tweezers, although widely used for single-cell manipulation, face several challenges. These include potential damage to the target, restriction to static environments, complex excitation setups, and interference outside the target area. To address these issues, we propose an ``invisible hydrodynamic tweezer'' utilizing near-zero index hydrodynamic metamaterials. This metamaterial-based device creates an equipotential resistance zone, effectively immobilizing particles in flowing fluids without disturbing the external flow field and without causing damage to the targets. Unlike traditional active control methods, our tweezer passively captures and releases particles by adjusting the flow channel, eliminating the need for continuous and stable excitation devices, thereby significantly simplifying the setup complexity. Furthermore, these tweezers can be modularly designed in different sizes to flexibly accommodate various application needs. Simulations and experimental validations demonstrated the non-interfering, stable trapping, and precise movement capabilities of these tweezers. This proposed technique holds significant potential for applications in biomedicine, microfluidics, and environmental monitoring.

\end{abstract}

\maketitle

 \section{Introduction}
Manipulating individual cells or tissue clusters is of significant importance in life science research, particularly in stem cell research~\cite{stem cell research}, regenerative medicine~\cite{regenerative medicine}, organ-on-a-chip technology~\cite{organ-on-a-chip technolog}, and tissue engineering~\cite{tissue engineering}. Since Ashkin first captured and manipulated bacteria and living cells in 1987~\cite{Ashkin-1,Ashkin-2}, various contactless tweezers have been widely applied in these fields, contributing to breakthroughs in microbiology, molecular biology, biophysics, and bioanalytical chemistry. Contactless tweezers utilize various physical forces to capture and manipulate particles. The most well-known optical tweezers~\cite{optical tweezers-1,optical tweezers-2,optical tweezers-3,optical tweezers-4} use the radiation pressure and gradient forces of a laser beam to create an optical potential well, capturing and manipulating microscopic particles. Magnetic tweezers~\cite{Magnetic tweezers} employ magnetic fields to exert forces on magnetic particles, precisely controlling their position and movement. Acoustic tweezers~\cite{Acoustic tweezers-1,Acoustic tweezers-2} use the acoustic radiation force generated by sound waves propagating through a medium, achieving precise particle manipulation by adjusting the phase and intensity of the sound waves. Other types include optoelectronic tweezers~\cite{optoelectronic tweezers}, which utilize dielectrophoresis effects on photoconductive materials to manipulate particles, and plasmonic tweezers~\cite{plasmonic tweezers}, which use locally enhanced electromagnetic fields generated by surface plasmon resonance to capture particles.

Despite the significant advancements in the development of various types of tweezers, several limitations persist. First, regarding the manipulated objects, some tweezers may cause damage during use. For instance, the high-power lasers in optical tweezers can inflict thermal damage on cells, while the enhanced electromagnetic fields in plasmonic tweezers can lead to localized heating, potentially harming biological tissues. Furthermore, many tweezers are limited to manipulating specific types of particles; for example, magnetic tweezers are restricted to magnetic particles, and optoelectronic tweezers depend on the polarization properties of the objects. Second, concerning the background environment, existing contactless tweezers cannot eliminate interference outside their functional area, which may lead to unintended disruptions. When multiple tweezers are used simultaneously, they can interfere with each other, reducing their precision. Moreover, these tweezers struggle to stably capture targets in flowing fluids. Third, the operational principles of current contactless tweezers require active control, necessitating a continuous and stable excitation source to maintain particle capture. This requirement renders the devices and platforms for these tweezers complex and energy-intensive. Although hydrodynamic tweezers~\cite{hydrodynamic tweezers-1,hydrodynamic tweezers-2}, which utilize fluid forces to manipulate particles, have been proposed to simplify equipment, they also demand high energy input and significantly impact areas outside their functional region. Recently, the advent of hydrodynamic metamaterials~\cite{hydrodynamic metamaterials-1,hydrodynamic metamaterials-2,hydrodynamic metamaterials-3,hydrodynamic metamaterials-4,hydrodynamic metamaterials-5,hydrodynamic metamaterials-6,hydrodynamic metamaterials-7,hydrodynamic metamaterials-8} offers a promising solution to several existing challenges. These metamaterials have introduced a new paradigm in flow field control, enabling precise manipulation of fluids through the meticulous design of artificial structures. This approach not only enhances the flexibility and accuracy of fluid manipulation but also facilitates the development of unprecedented functional devices. For example, a hydrodynamic cloak~\cite{hydrodynamic cloak-1,hydrodynamic cloak-2,hydrodynamic cloak-3,hydrodynamic cloak-4,hydrodynamic cloak-5,hydrodynamic cloak-6,hydrodynamic cloak-7,hydrodynamic cloak-8} can reroute fluid around a device and restore it to its original path, making the flow field appear unchanged regardless of the cloak’s presence, thus achieving hydrodynamic cloaking. Furthermore, Boyko et al. introduced the concept of hydrodynamic shielding~\cite{hydrodynamic shield-1,hydrodynamic shield-2}, which creates a uniform pressure distribution in the functional area, eliminating the pressure differential forces exerted by the fluid on the object. These two functionalities inspire the design of a new class of hydrodynamic tweezers. However, to date, research has yet to simultaneously achieve both hydrodynamic shielding and cloaking, and the principles of metamaterials have not yet been applied to the design of contactless tweezers, which aim to create a damage-free, non-interfering, and passive universal tool for controlling particles in flowing fluids.

\begin{figure}[htp]
	\centering
	\includegraphics[width=\linewidth]{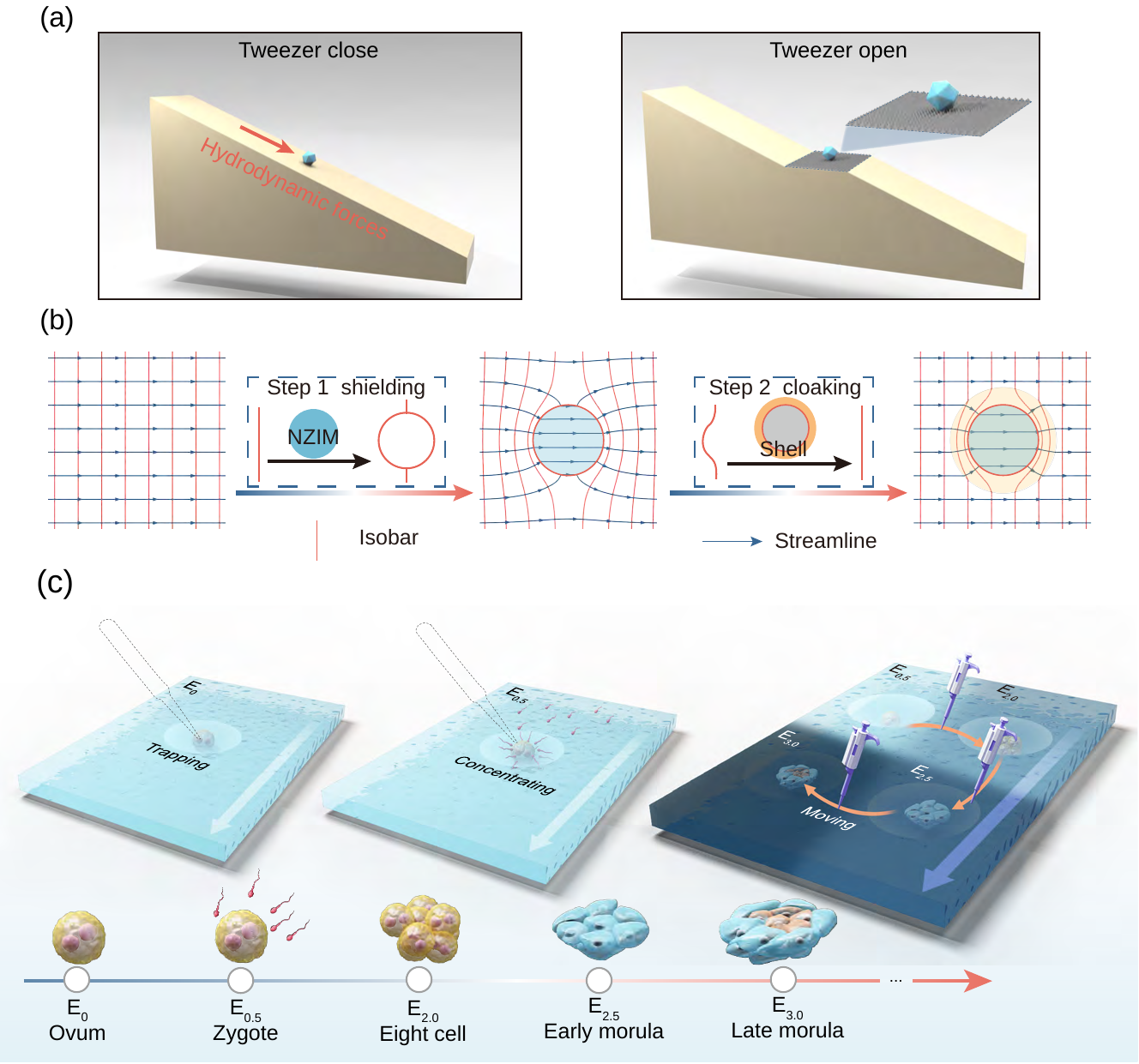}\\
	\caption{Design and Application of Invisible Hydrodynamic Tweezers. \textbf{a}. Traditional potential well-based tweezers create a central force field to trap particles, and the proposed invisible hydrodynamic tweezers utilize an equipotential viscous region to trap particles. \textbf{b}. Design process of the invisible hydrodynamic tweezer. \textbf{c}. Application of the hydrodynamic tweezers in in vitro embryo culture, demonstrating trapping, concentrating, and moving capabilities for different developmental stages from ovum to late morula, showcasing their potential in bioengineering.}
	\label{F1}
\end{figure}

In this study, we propose an invisible hydrodynamic tweezer capable of precisely decelerating and trapping particles moving within a flow field, as well as controlling their movement. The proposed tweezers are based on a novel principle for manipulating particles. Instead of using a potential well to trap particles, as shown in Figure~\ref{F1}\textbf{a}, these tweezers create an equipotential viscous region to immobilize the particles. This principle imposes no specific requirements on the particles being manipulated, nor does it exert any additional force or cause damage to them. Moreover, our hydrodynamic tweezers are ``invisible'' to the external environment, meaning they do not disturb the flow field outside the tweezers. The design concept of the invisible hydrodynamic tweezer integrates the principles of shielding and cloaking from hydrodynamic metamaterials. First, drawing on analogies from photonics and thermodynamics, we introduce the concept of hydrodynamic zero-index materials, using them as the core component of the tweezer to capture particles in the flow field. Then, by applying the principle of scattering cancellation, we design the outer layer of the tweezer to eliminate its interference with the external flow field, as shown in Figure~\ref{F1}\textbf{b}. The above process can be passively achieved solely through the design of the fluid channel geometry. Once the particles are captured, they can remain fixed without the need for continuous energy input. Experiments demonstrated the tweezers' ability to concentrate and capture particles. By using elastic materials, we achieved the mobility and precise positioning of particles with the tweezers. These functionalities have substantial applications in the life sciences. For example, during in vitro embryo culture, the tweezers can immobilize an ovum and concentrate sperm around it. Additionally, at various stages of embryo development, tissues can be transferred to different culture areas in a non-contact and non-destructive manner, as shown in Figure~\ref{F1}\textbf{c}. These features make the invisible hydrodynamic tweezer a promising tool for bioengineering applications.

 \section{Theory}
 \subsection{The governing equations in hydrodynamics}
The Poiseuille model serves as a fundamental theory for understanding fluid dynamics in confined spaces, such as pipes and narrow channels. Its applications are extensive across both theoretical fluid mechanics and practical industrial processes. This model is instrumental in modeling the flow of blood through vessels, designing and evaluating fluid movements within reactors, simulating the seepage of groundwater through the porous media of soil and rock, and enhancing fluid transport efficiency in microchannels. Specifically, for Poiseuille flow occurring between two parallel plates, the velocity distribution is governed by the equation\cite{Poiseuille}:
\begin{equation}
	\mathbf{v}_{\Vert} = -\frac{1}{2\mu}z(z-h)\nabla_{\Vert} p.
	\label{E1.1}
\end{equation}
Here, $\mathbf{v}_{\Vert}$ represents the fluid's velocity parallel to the plates, $\mu$ is the viscosity coefficient of the fluid, crucial for determining resistance to flow. The parameter $h$ refers to the height of the fluid channel, while $z$ is the coordinate perpendicular to these plates. The operator $\nabla_{\Vert}$ signifies the two-dimensional gradient along the $x-y$ plane, and $P$ represents the pressure exerted within the fluid. By integrating the velocity profile $\mathbf{v}_{\Vert}(z)$ over the entire gap between the plates, we can derive the volumetric flow rate $Q$, which quantifies the total volume of fluid flowing between the plates per unit time:
\begin{equation}
	Q = \int_{0}^{h} \mathbf{v}_{\Vert}(z) dz = \frac{h^3}{12\mu}\nabla_{\Vert} p.
	\label{E1.2}
\end{equation}
Considering the conservation of flow for an incompressible fluid flowing between parallel plates, where $\nabla Q = 0$, the following Laplacian form of the flow equation can be obtained:
\begin{equation}
	\frac{h^3}{12\mu}\nabla^2_{\Vert} P = 0.
	\label{E1.3}
\end{equation}
 
 \subsection{Hydrodynamic near-zero index metamaterial}
\begin{figure}[htp]
	\centering
	\includegraphics[width=\linewidth]{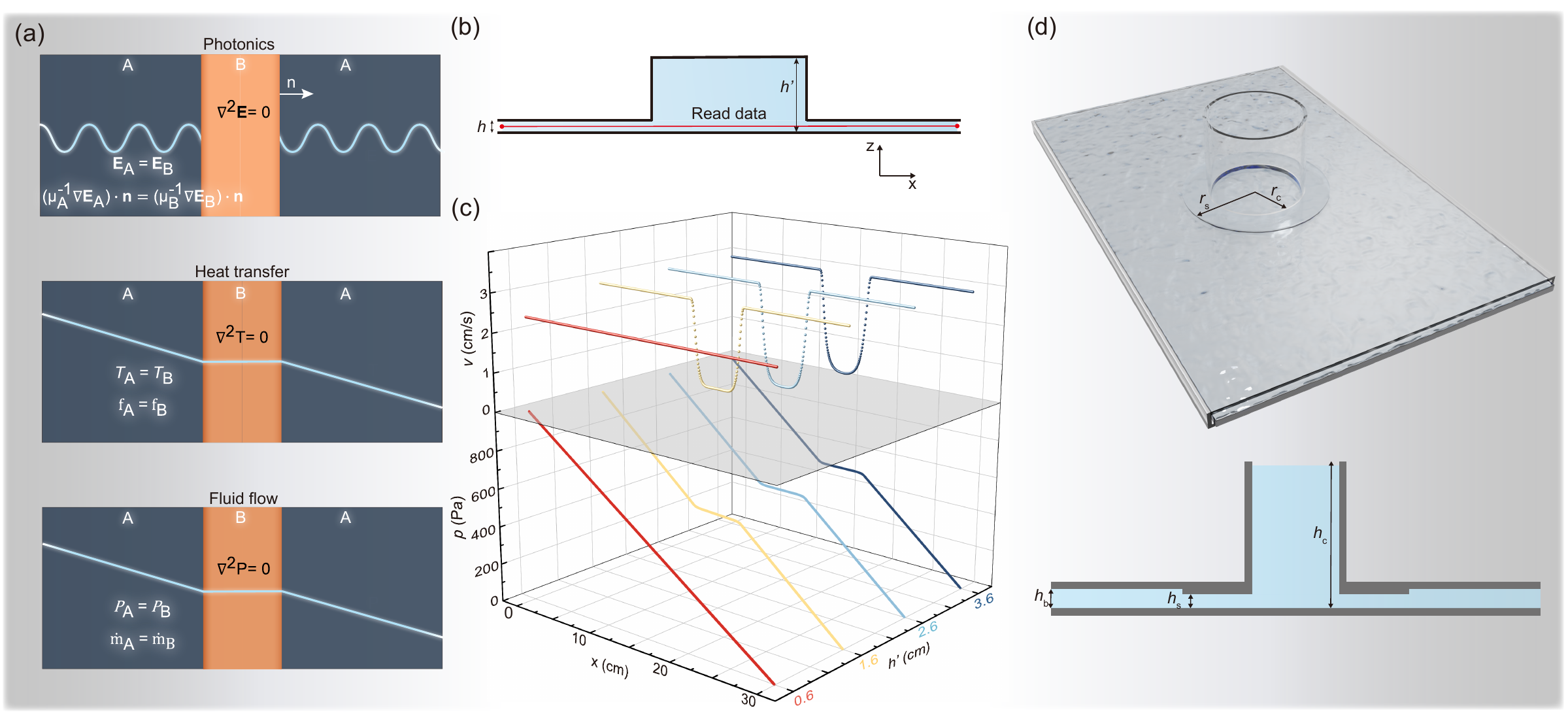}\\
	\caption{\textbf{a}. Zero-index phenomenon in different physical fields: the blue region represents ordinary materials, while the central orange region represents near-zero index materials. \textbf{b}. Achieving near-zero index effect in fluid flow using an expansion structure in the planar flow model. \textbf{c}. Near-zero index effects achieved with different heights of the expansion structure. The higher the $h_{\rm{c}}$, the closer the fluid velocity approaches zero in this region, and the pressure gradient tends to zero. \textbf{d}. Three-dimensional structure of the hydrodynamic tweezer.}
	\label{F2}
\end{figure}

 The concept of zero index metamaterials first emerged in the field of photonics~\cite{zero photonics-1,zero photonics-2}, distinguished by their remarkable properties of infinite phase velocity and wavelength. By engineering artificial microstructures, it is possible to create near-zero index metamaterials (NZIM), which allow waves to exhibit a uniform phase distribution across the material. This characteristic is particularly difficult to replicate in natural materials. Such unique behavior is demonstrated in Figure~\ref{F2}\textbf{a}, highlighting the advanced capabilities of NZIMs.
 
 The concept of NZIM extends beyond wave systems to include diffusion systems as well~\cite{zero thermal}. For example, in the context of heat transfer, an NZIM can be conceptualized as a material possessing infinite thermal conductivity, which enables heat to traverse through it without any reduction in temperature. This property can be analogously linked to the steady-state Fourier heat conduction law, expressed as $\kappa \nabla^2 T = 0$. When compared with Equation~(\ref{E1.3}), both exhibit remarkably similar forms, where $T$ in the heat conduction equation corresponds to the $P$ in the fluid dynamics equation. Thus, we characterize a fluid's near-zero index metamaterial (NZIM) as a material through which, when fluid flows, the pressure variation is minimal, denoted as $\nabla p \rightarrow 0$. Additionally, from Equation~(\ref{E1.3}), we observe that the term $h^3/12 \mu$ is analogous to $\kappa$ in heat conduction systems. Therefore, while theoretically achieving a zero-index condition necessitates the parameter $h^3/12 \mu$ approaching infinity, this scenario is impractical in real-world applications. Firstly, when $h$ becomes exceedingly large, the assumptions underlying the Poiseuille model no longer hold, rendering Equation~(\ref{E1.3}) inapplicable. Secondly, significantly altering the fluid's viscosity coefficient $\mu$ to near zero is exceedingly challenging. Consequently, alternative strategies must be explored to realize effects similar to those of NZIMs.
 
 We propose employing channels with abrupt expansions to simulate NZIM-like effects. As depicted in Figure~\ref{F2}\textbf{b}, the channel has an initial thickness of $h$ and expands into a region that resembles a well, allowing the fluid to ascend in the $z$-direction to a height of $h^{\prime}$. According to the continuity equation, the volumetric flow rates before and after the expansion are equal, represented as $Q = \langle \mathbf{v}_{\Vert} \rangle h = \langle \mathbf{v}_{\Vert}^{\prime} \rangle h^{\prime} = Q^{\prime}$. Here, $\langle \mathbf{v}_{\Vert} \rangle$ and $\langle \mathbf{v}_{\Vert}^{\prime} \rangle$ denote the depth-averaged velocities before and after the expansion, respectively. Consequently, upon entering the expanded region, the average velocity of the fluid reduces to $\langle \mathbf{v}_{\Vert} \rangle h/h^{\prime}$, illustrating a decrease in flow velocity due to the increased cross-sectional area. In an ideal scenario, if $h^{\prime}$ were to approach infinity, $\langle \mathbf{v}_{\Vert}^{\prime} \rangle$ would tend towards zero. However, in the presence of gravity, $h^{\prime}$ remains finite, with its maximum value dictated by the static pressure equilibrium of the fluid. As $\langle \mathbf{v}_{\Vert}^{\prime} \rangle$ approaches zero, the pressure loss due to viscosity diminishes significantly, becoming virtually negligible. Consequently, this scenario mirrors conditions akin to an ideal fluid, where pressure remains constant along any given streamline at the same elevation. This setup closely approximates the zero-index objective, where $\nabla p \rightarrow 0$. We explore this concept in greater depth in Supplementary Note 1.
 
We conducted a simulation to corroborate the analysis presented, setting a gap of 0.6 cm between two plates and using $95\%$ glycerin as the fluid under a total pressure difference of $\Delta p = 1000~\rm{Pa}$. Figure~\ref{F2}\textbf{c} displays the pressure and velocity distributions across various heights of the expanded region. The line chart demonstrates that as the height $h_c$ of the expanded region increases, the fluid's velocity significantly decreases upon entering the central expansion, while the pressure gradient rapidly approaches zero. This results in the attainment of the near-zero index effect, aligning closely with our theoretical predictions.

 \subsection{Designing invisible hydrodynamic tweezers based on near-zero index metamaterials}
In fluid dynamics, an object immersed in a fluid encounters several forces. The pressure drag force results from the differential pressure across the front and back of the object, propelling it in the direction of the flow. Viscous friction force arises due to the fluid's viscosity, influenced by the relative motion between the fluid and the object. Inertial forces come into play when the object accelerates or decelerates, with the fluid's inertia exerting an impact. Gravity acts downward, representing the object's weight, which is often analyzed in conjunction with the buoyant force that opposes it. Additionally, in large-scale rotational systems, such as on Earth, the Coriolis force becomes significant, affecting the motion of objects in these rotating frames. In this work, our focus is on the behavior of small objects in low-speed, viscous laminar flow conditions, where the predominant forces are the pressure differential force and the viscous friction force. Typically, objects in such environments move in accordance with these forces, flowing along the direction of the fluid. To effectively trap such a moving object, a region of uniform pressure is necessary to neutralize the pressure differential force. Simultaneously, the fluid velocity within this region must be close to zero to ``brake'' the object through viscous friction. Remarkably, the hydrodynamic near-zero index region we previously introduced meets these criteria well, rendering it an excellent choice for the main mechanism of the hydrodynamic tweezer.

However, if the tweezer is comprised solely of the NZIM region, it could inevitably influence the surrounding area, such as altering the fluid flow direction. This issue is not unique to hydrodynamic tweezers but also affects other types when trapping particles within a fluid. This can become particularly problematic in scenarios where tweezers are densely deployed on a large scale to trap different particles, as the interference from one set of tweezers can disrupt the trapping capabilities of adjacent ones. While addressing this issue is challenging for tweezers that utilize optical, magnetic, or acoustic forces, hydrodynamic tweezers offer a unique solution. By employing the concept of hydrodynamic cloaking, we can effectively mitigate such interference. The design process for the invisible hydrodynamic tweezer is outlined in Figure~\ref{F1}\textbf{b}. Initially, the NZIM region forms the core of the tweezer, eliminating internal pressure differentials and ensuring minimal fluid velocity. Subsequently, an outer shell layer is designed to eliminate any external environmental interference caused by the tweezer.

In practice, this process can be achieved by adjusting the inter-layer heights. The structure of the designed tweezer is shown in Figure~\ref{F2}\textbf{d}, where $r_{\rm{c}}$ and $r_{\rm{s}}$ represent the radii of the core region and the shell layer, respectively. The inter-layer heights of the fluid in these regions are denoted as $h_{\rm{c}}$ and $h_{\rm{s}}$, while $h_{\rm{b}}$ represents the inter-layer height in the background region. In the depicted flow channel, fluid flows uniformly from one end, moves along the horizontal plane, and exits uniformly from the other end. At this point, the pressure gradient along the flow direction is linear, and the interfaces between different regions need to satisfy equal pressure and mass conservation. By applying these conditions to the general solution of Equation~(\ref{E1.3}) in polar coordinates, we can determine the necessary geometric relationship for the inter-layer height $h_{\rm{s}}$ to maintain these balances:
\begin{equation}
	h_{\rm{s}} = \sqrt[3]{\frac{r_{\rm{s}}^2 - r_{\rm{c}}^2}{r_{\rm{s}}^2 + r_{\rm{c}}^2}h_{\rm{b}}^3}.
	\label{E3.1}
\end{equation}
Please see Supplementary Note 2 for details. When considering the influence of gravity, the pressure $P$ in Equation~(\ref{E1.3}) should be modified to $p + \rho g h$. However, since $\nabla_{\Vert} (\rho g h) = 0$ on the same horizontal plane, this term can be ignored. To achieve a near-zero index effect, the core region's thickness $h_{\rm{c}}$ must be substantially greater than that of the background. It is worth noting that Equation~(\ref{E3.1}) is derived under ideal boundary conditions, where $\nabla_{\Vert} p = 0$ in the core region. In practical scenarios, as depicted in Figure~\ref{F2}\textbf{c}, there exists a minimal smooth transition zone between the background and the NZIM region where the velocity and pressure gradients are not zero, which is not accounted for in our initial design. Therefore, after deriving the geometric parameters of the channel structure from Equation~(\ref{E3.1}), further optimizations and adjustments in simulations and experiments are necessary to refine the design and enhance its functionality.

 \section{Numerical demonstration of invisible hydrodynamic tweezers}
We confirm the efficacy of the hydrodynamic tweezers through simulations in COMSOL Multiphysics. The dimensions of the background region are 30~cm in length, 20~cm in width, and 6~mm in height.The height of the flow channel is 6 mm, which is much smaller than the horizontal scale, meeting the requirements of the Poiseuille model. The core region, or the hydrodynamic zero-index area, has a radius of $r_{\rm{c}} = 3~\rm{cm}$. To achieve a near-zero index and effectively trap particles, the height of this core region must substantially exceed that of the surrounding background. The impact of this height variation on the tweezer's trapping efficiency is detailed in Supplementary Note 3. In our simulations, we selected the maximum height achievable by the liquid surface, $h_{\rm{c}} = 5~\rm{cm}$, which directly correlates with the pressure boundary conditions at both the inlet and outlet, aiming to maximize performance. The external diameter of the shell region is $r_{\rm{s}} = 4.5~\rm{cm}$, and as dictated by Equation~(\ref{E3.1}), the corresponding height $h_{\rm{s}}$ should be 4.36 mm. Based on previous analyses, minor adjustments to $h_{\rm{s}}$ are necessary to optimize performance. We anticipate that the optimal $h_{\rm{s}}$ will be slightly greater than 4.36~mm; thus, parameterized scanning is employed to identify an appropriate height close to this value. Post-optimization, we have set $h_{\rm{s}} = 4.79~\rm{mm}$ as the parameter for our simulation settings, as detailed in Supplementary Note 4. The model's inlet features a constant pressure boundary condition of $P$ = 1000~Pa, and the outlet has a boundary condition of $P$ = 0~Pa. All other walls are subject to no-slip boundary conditions. The fluid's properties correspond to those of a 95\% concentrated glycerin solution ($\mu = 0.63~\rm{Pa\cdot s}$).

\begin{figure}[htp]
	\centering
	\includegraphics[width=\linewidth]{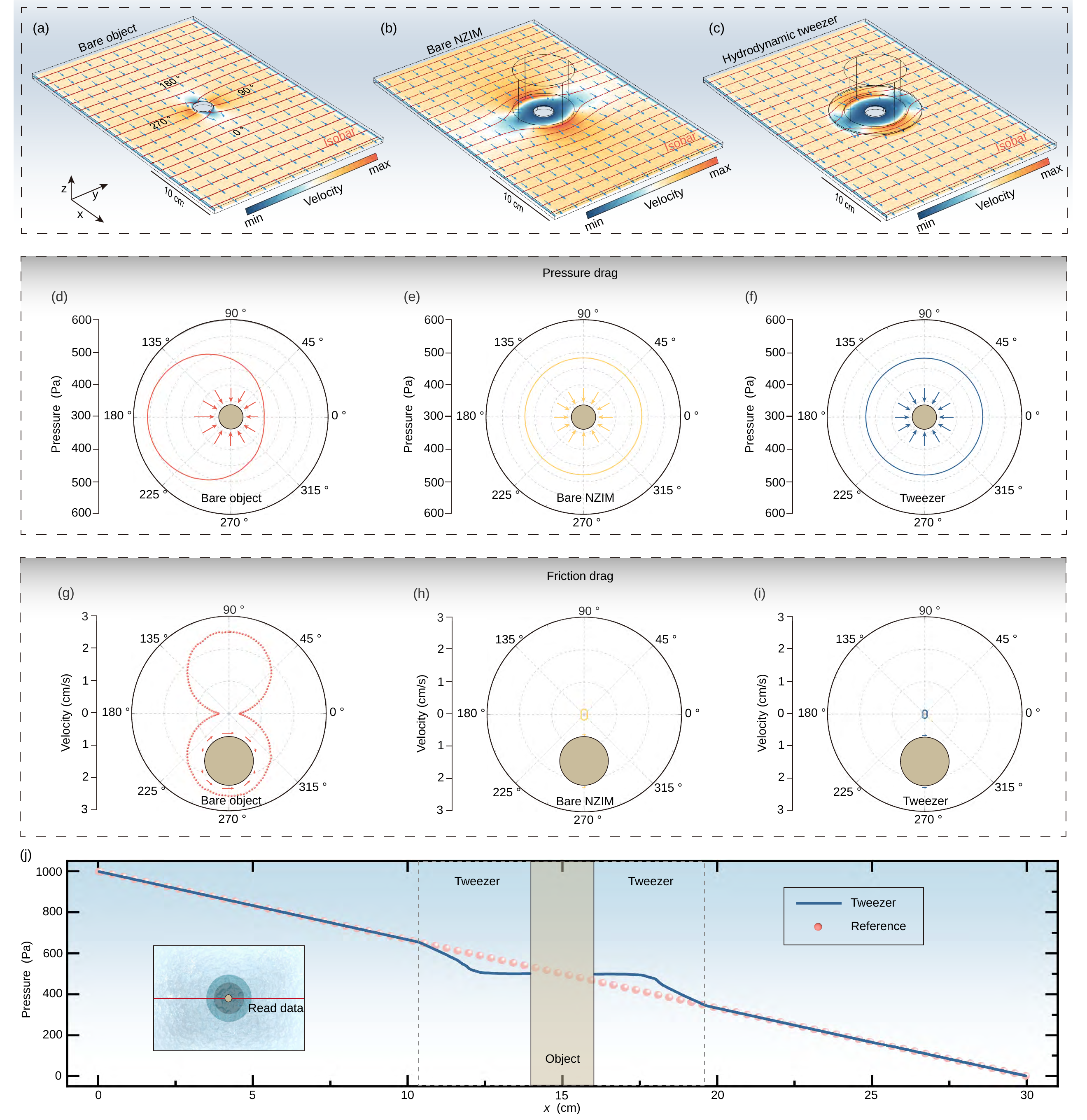}\\
	\caption{Steady-state flow field profiles ($z=3~\rm{mm}$ plane). \textbf{a}. Velocity and pressure distribution when only an object is present in the channel. \textbf{b}. A NZIM region is present in the flow field, the velocity within the region tends to zero, and the pressure is uniform. \textbf{c}. A invisible hydrodynamic tweezer does not disturb the external flow field distribution and features regions of uniform pressure within the device. \textbf{d} - \textbf{f}. Pressure distribution around the object under different conditions. The data selected in the figure are centered on the object, at a radius of 1.1 cm. \textbf{g} - \textbf{i}. Velocity distribution of the fluid around the object under different conditions. The data also come from a circle with a radius of 1.1 cm. \textbf{j}. Comparison of the pressure distribution in the flow field when using the invisible hydrodynamic tweezer with the reference group, which is the flow field with no obstacles.}
	\label{F3}
\end{figure}

First, we examine the forces acting on a large cylindrical object located at the center of the flow field, with a radius of $1~\rm{cm}$ and a height of $0.4~\rm{cm}$. Notably, the object's shape is irrelevant to the analytical outcomes, as corroborated in Supplementary Note 5, where simulations involving various shapes demonstrated no effect on device functionality. The steady-state velocity field and pressure distribution are displayed in Figures~\ref{F3}\textbf{a} - \textbf{c}, depicting scenarios with the bare object, bare NZIM, and a zero-index cloak, respectively. Data captured from the $x-y$ plane ($\rm{z~=3~mm}$) are presented here, with results from other positions and angles detailed in Supplementary Notes 6 and 7. The introduction of an object into the fluid flow scatters the surrounding flow, altering fluid velocity and disrupting pressure distribution. Encapsulating the object in a zero-index metamaterial region, characterized by nearly zero velocity and a uniform pressure field, significantly reduces scattering effects on the nearby flow. This setup also suggests interaction between the NZIM region and the ambient flow. To counteract the effects of the NZIM on the background flow, the hydrodynamic tweezer incorporates an additional structural layer, which not only suppresses interactions between the object and the flow but also neutralizes the tweezer's influence on the external environment.

In the following analysis, we examine the hydrodynamic forces exerted on the object under varying conditions, initially focusing on the pressure drag. This drag is primarily due to the non-uniform pressure distribution across the object's surface. As illustrated in Figure~\ref{F3}\textbf{d}, positioning the object directly in the flow field leads to higher pressure at the front, which generates a force in the direction of the flow. Conversely, as depicted in Figures~\ref{F3}\textbf{e} and \textbf{f}, when the pressure distribution is even around the object, the forces from all directions negate each other, resulting in no pressure drag.

Also, the object is subject to friction drag, which results from the relative motion between the fluid and the object. The intensity of this force depends on the fluid's viscosity and the velocity of the flow. In simple terms, higher flow velocities lead to increased friction drag; conversely, as the flow velocity nears zero, the friction drag similarly diminishes. Figures~\ref{F3}\textbf{g} - \textbf{i} display the velocity distribution surrounding the object. The data reveals that the friction drag is minimal in the direction of the flow and maximal on the object's sides. Within the near-zero index metamaterial (NZIM) region, where the fluid velocities around the object are close to zero, the object experiences negligible friction drag.

We further assessed the influence of the hydrodynamic tweezer on the flow field external to the target area. As depicted in Figure~\ref{F3}\textbf{j}, a line encompassing both the background and the tweezer was chosen for this analysis. We compared the pressure data along this line to a control scenario—namely, a flow field devoid of any objects. The findings show that the pressure in the background region remains consistent across both scenarios, indicating that the tweezer does not disturb the flow field outside its operational area. Moreover, the pressure within the core region of the tweezer is stable, confirming that objects within this zone are not subjected to pressure differential force.

\begin{figure}[htp]
	\centering
	\includegraphics[width=\linewidth]{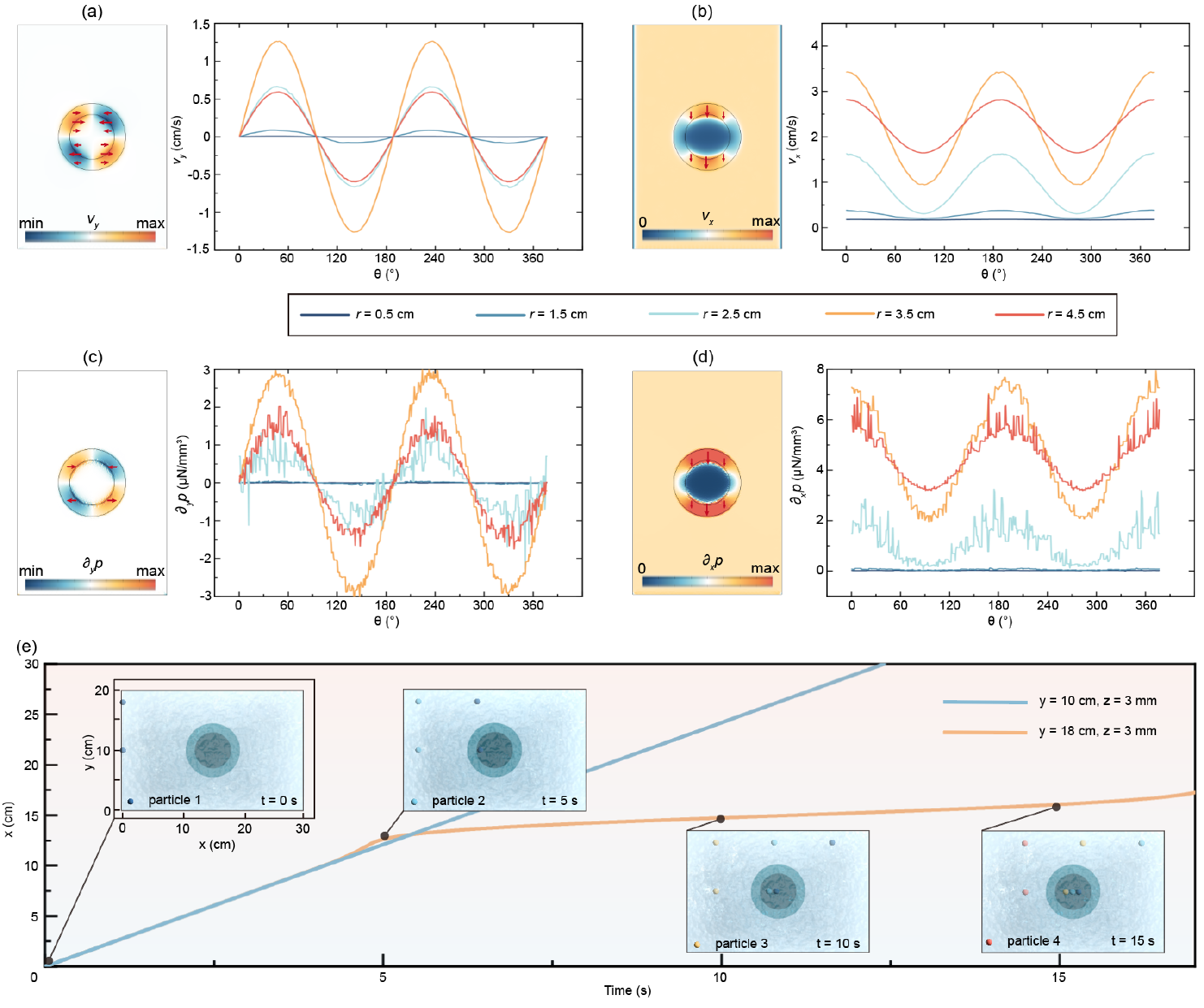}\\
	\caption{\textbf{a} and \textbf{b}. Velocity distributions in the y and x directions at different positions within the hydrodynamic tweezers. The data is collected from circles centered on the tweezers with radii of 0.5 cm, 1.5 cm, 2.5 cm, 3.5 cm, and 4.5 cm. \textbf{c} and \textbf{d}. Pressure gradient distributions inside the hydrodynamic tweezers. \textbf{e}. The demonstration of tweezers functionality. The line graph displays the displacement of two particles released simultaneously at the entrance over time, with particles passing through the zero-index cloak being retained in the central region. The insets show the position plots of particles released every four seconds at the entrance at different times.}
	\label{F4}
\end{figure}

Next, we delve into the mechanism through which the tweezer manipulates small particles. For particles of very small volume, the impact of pressure drag is negligible, leaving friction drag as the dominant influencing force. Figures~\ref{F4}\textbf{a} and \textbf{b} illustrate the distribution of velocity components $v_y$ and $v_x$ at various positions within the tweezer. Particularly for the $y$-direction velocity component, the fluid velocity at the front of the tweezer (where $0^{\circ}~\textless~\theta~\textless~180^{\circ}$) is directed towards the center, peaking at the positions of $45^{\circ}$ and $135^{\circ}$. In the radial direction, the velocity nears zero close to the core of the tweezer. As a result, particles from the inflow direction are drawn towards the tweezer's center by viscous forces as they traverse its field. For the $x$-direction velocity component, the speed peaks in the direction of incoming flow ($90^{\circ},270^{\circ}$) and is minimized at the tweezer's lateral sides ($0^{\circ}, 180^{\circ}$). Likewise, velocities near the tweezer's interior approach zero. Consequently, particles moving through the tweezer undergo acceleration in the $x$ direction initially, and then decelerate to a halt as they reach the central region. Given the diminutive size of the particles, the pressure drag they encounter can be estimated by multiplying their volume with the pressure gradient at their location. Figures~\ref{F4}\textbf{c} and \textbf{d} display the pressure gradients in the $y$ and $x$ directions at various positions, respectively. Within the tweezer, the directions of the pressure drag and friction drag are generally aligned, and they both rapidly diminish to zero in the core region.

The ability of the tweezer to immobilize particles can be effectively demonstrated using the particle tracing module in COMSOL. The line graph depicted in Figure~\ref{F4}\textbf{j} shows the displacement over time of particles as they pass through the hydrodynamic tweezer in comparison to those that do not. Initially, the displacement of both sets of particles is identical, which suggests that the tweezer does not impact the external flow field. However, as the observation continues, particles moving through the tweezer exhibit little to no displacement change, while those outside swiftly move towards the exit. If particles are continuously introduced at the entrance, the following pattern emerges: particles that outside the tweezer exit at a steady speed, whereas those within the tweezer accumulate in the central area, as depicted in the four insets of the graph. 

In the analyses presented above, all scenarios were conducted under a linear pressure gradient. The effectiveness of the tweezers remains excellent even in cases with a nonlinear pressure gradient, as detailed in Supplementary Note 8. In Supplementary Note 9, we simulated scenarios involving multiple tweezers used simultaneously and demonstrated that they do not interfere with each other. As one of the important dimensionless parameters in fluid dynamics, the Reynolds number effectively reflects the flow state of fluids. In our simulations, the model's Reynolds number was approximately 0.28, indicating typical laminar flow. The tweezers' performance remained stable even in higher Reynolds number flow scenarios, as detailed in Supplementary Note 10. Additionally, in Supplementary Note 11, we verified the tweezers' functionality across different scenarios and sizes. These simulations collectively prove that the proposed hydrodynamic tweezers exhibit robust performance across various application scenarios. Finally, we conducted a mesh independence analysis, confirming the reliability of the conclusions presented above (see Supplementary Note 12).

\section{Experimental demonstration of invisible hydrodynamic tweezers}
\begin{figure}[htp]
	\centering
	\includegraphics[width=0.8\linewidth]{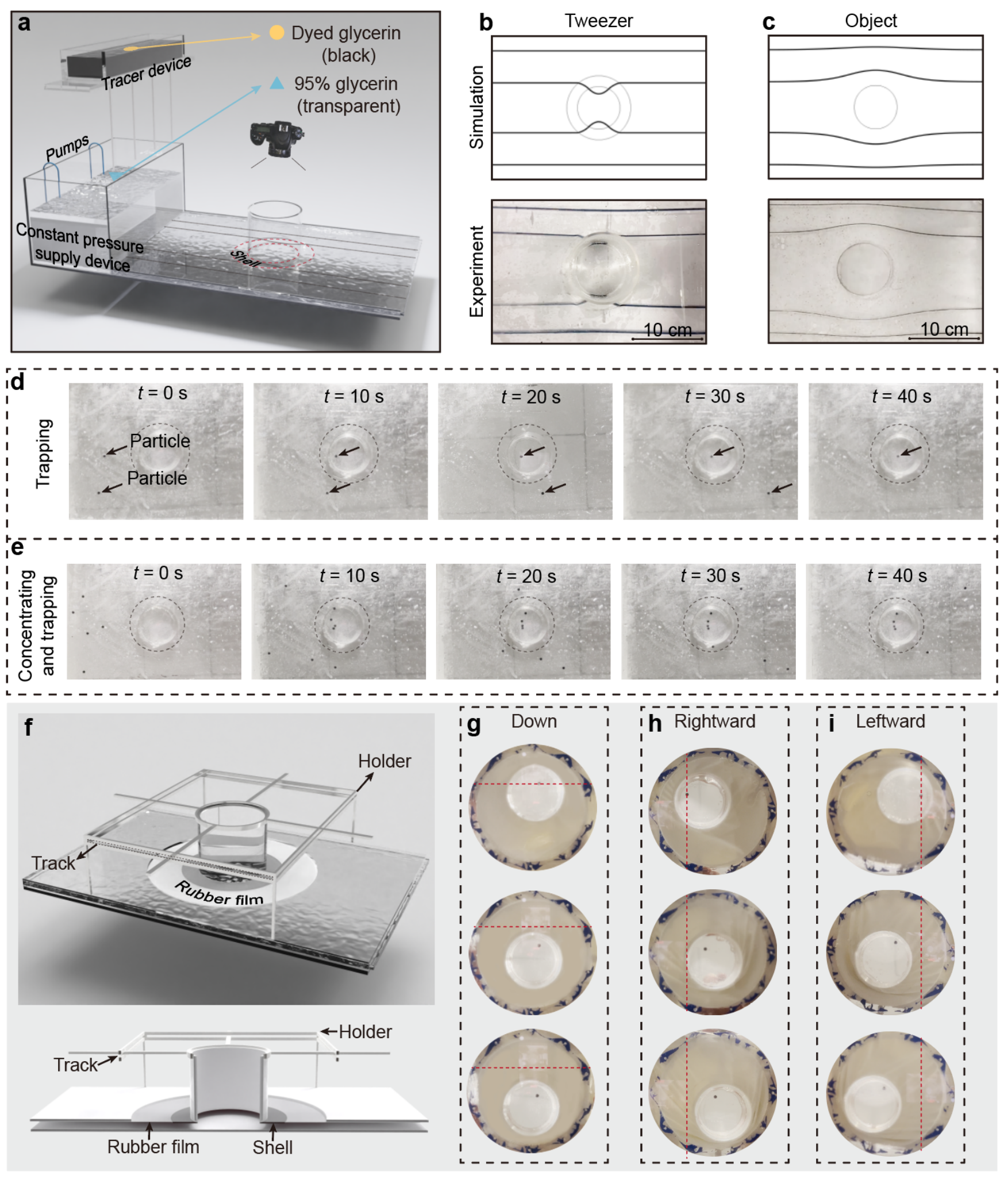}\\
	\caption{Experimental verification of the proposed invisible hydrodynamic tweezer. \textbf{a}. Schematic diagram of the experimental setup for a fixed-position tweezer. The entire setup is constructed using transparent acrylic sheets to facilitate observation. A tracing device is used to visualize the streamlines. \textbf{b} and \textbf{c}. Distribution of streamlines in two different scenarios along with corresponding simulation results, demonstrating that the invisible hydrodynamic tweezer does not affect the flow field outside the tweezer. \textbf{d}. Demonstration of the tweezers' ability to capture a single silicon nitride particle. Particles passing through the tweezer are stably trapped within it, while those outside remain unaffected. \textbf{e}. Demonstration of the tweezers' capability to gather and fix multiple particles within the flow field. \textbf{f}. Schematic diagram of the experimental setup for a movable tweezer. In this setup, part of the acrylic board is replaced with a rubber film to ensure the tweezer's mobility. The fixed and track devices maintain the horizontal position of the tweezer. A cross-sectional view illustrates the positions of each component. \textbf{g}, \textbf{h} and \textbf{i}. Demonstrations of the tweezer moving particles downward, rightward, and leftward, respectively.}
	\label{F5}
\end{figure}

We conducted experiments to validate the functionality of the invisible hydrodynamic tweezers and proposed two practical methods for using these tweezers to manipulate particle movement. First, we examined the tweezers' ability to capture particles at a fixed position. The experimental setup, as shown in Figure~\ref{F5}\textbf{a}, consists of three main components. The flow channel, made of transparent acrylic sheets, forms a passage that is 30 cm long, 20 cm wide, and 6 mm thick. The hydrodynamic tweezers are positioned at the center of this channel to capture particles passing through this location. The setup includes a constant pressure supply system, implemented using an acrylic jar equipped with pumps. The bottom of the jar connects to the main body, allowing liquid to flow into the channel from the inlet under gravity. As long as the pump maintains a constant liquid level in the jar, the pressure at the inlet remains stable, fulfilling the required boundary conditions. Additionally, the experimental setup features a tracer device filled with colored liquid. During the experiments, this colored liquid is released from four thin tubes of the tracer device, effectively illustrating the streamlines. The experiment uses 95$\%$ glycerol due to its high viscosity and transparency, which helps maintain stable experimental conditions. To prevent interference from the tracer's colored liquid, we used the same 95$\%$ glycerol base but added pigment, ensuring it does not alter the original flow dynamics.

Figure~\ref{F4}\textbf{b} illustrates the arrangement of four distinct streamlines in the flow field when using the hydrodynamic tweezers. The positions of the streamlines were carefully selected: two pass through the tweezers, while the other two bypass them, enabling a comprehensive analysis of the streamline distribution across different regions. The results show that the tweezers maintain the integrity of the flow field, with streamlines remaining straight outside the tweezers, as if they were not present. Within the tweezers, however, the streamlines converge towards the center, consistent with the simulation results. This observation confirms that the tweezers can directly adjust the internal flow distribution without disrupting the external flow, demonstrating the "invisibility function" of the hydrodynamic tweezers, a feature unattainable by other contactless tweezers. For comparison, Figure~\ref{F4}\textbf{c} presents a control scenario with a cylindrical object of 3 cm in radius and 6 mm in height in the flow field, showing that the presence of an ordinary object disturbs the flow field, causing the streamlines to deviate.

Next, we validate the functionality of the tweezers, specifically their ability to capture particles moving within the fluid. Initially, we introduced two smooth silicon nitride particles with a diameter of 4 mm at different positions near the inlet and monitored their trajectories in the flow field, as illustrated in Figure~\ref{F5}\textbf{d}. The results show that particles not passing through the tweezers moved uniformly along straight paths, while those entering the tweezers were effectively immobilized within the device, consistent with the simulation predictions shown in Figure~\ref{F4}\textbf{e}. Subsequently, we randomly introduced several silicon nitride particles at the inlet. The outcomes, depicted in Figure~\ref{F5}\textbf{e}, demonstrate the tweezers' capability to gather and capture particles within the fluid, confirming their effectiveness in practical applications.

Generally, tweezers need to not only stably capture particles but also move them. Here, we demonstrate a simple method for moving particles using hydrodynamic tweezers, with the device structure illustrated in Figure~\ref{F5}\textbf{f}. This device replaces part of the acrylic plate with a highly elastic rubber film, which is connected to the hydrodynamic tweezers. This design allows the tweezers to move, thereby facilitating the movement of particles. The specific position of the rubber film can be seen in the cross-section view, where it should be flush with the rest of the channel. The acrylic plate is equipped with a set of fixtures and a track, through which the support of the tweezers passes, ensuring that the tweezers maintain a consistent horizontal level while moving.

Figures~\ref{F5}\textbf{g}, \textbf{h} and \textbf{i} show the tweezers moving particles downstream, to the right, and to the left, respectively. The tweezers are moved in the desired direction, causing the hydrodynamic forces within the tweezers to move the particles accordingly. By doing so, we successfully moved the particles several centimeters in the desired direction. Notably, while the tweezers can directly move particles downstream, they cannot move particles strictly right, left, or upstream. As shown in Figure~\ref{F4}, the hydrodynamic forces within the tweezers, including pressure drag and friction drag, are always greater than or equal to zero in the downstream direction. Therefore, without external excitation, tweezers relying solely on hydrodynamic forces cannot move particles upstream.

Theoretically, the rubber film should remain flat during the movement to maintain the channel's integrity. In practice, this can be achieved by pre-stretching the membrane during installation to ensure it remains flat upon contraction. Although our experimental setup did result in some wrinkles on the rubber membrane due to technical limitations, these wrinkles affected only the background flow field and did not impact the flow within the tweezers. Therefore, the functionality of the tweezers in moving particles remains unaffected, and our experimental conclusions are valid. Furthermore, in Supplementary Note 13, we propose an alternative method for moving the tweezers to circumvent the aforementioned issues. This approach allows for the movement of particles over longer distances without being constrained by the stretch limits of the material. However, it is more complex to manufacture and incurs higher costs.

In the aforementioned experiments, the hydrodynamic tweezers effectively captured and held particles without any external energy input, maintaining stability throughout the process. This passive mechanism is more energy-efficient compared to other actively controlled contactless tweezers in certain scenarios. Despite their simple structure, as shown in the real photos in Supplementary Note 14, the tweezers demonstrate significant effectiveness, thereby lowering the barriers for practical application and advancing the broader use of contactless tweezers.

\section{Discussion and conclusion}
The tweezers introduced in this study are not limited to direct applications in life sciences. They can be coupled with other physical fields to provide even more functionalities. For example, integrating this technology with heat transfer mechanisms could serve as a method for controlling convective heat transfer~\cite{multiphysics-1,multiphysics-2,multiphysics-3,multiphysics-4,multiphysics-5,multiphysics-6}. This cross-disciplinary potential significantly broadens the scope of applications, enhancing the utility and impact of invisible hydrodynamic tweezers. In addition, the concept of hydrodynamic zero-index materials proposed in this study, due to their mechanical protection properties, holds promise for a variety of applications (see Supplementary Note 15), such as shielding bridge piers from river flow impacts. Furthermore, the method of adjusting the channel height not only facilitates the realization of the hydrodynamic zero-index phenomenon but also offers a means of achieving arbitrary flow velocities. For instance, increasing the channel height can reduce the flow velocity in a specific region, while decreasing the height can increase it. This approach, when combined with hydrodynamic metamaterials, has the potential to create devices with additional functionalities, such as hydrodynamic concentrators. Finally, there is room for further improvement in this device. Currently, we achieve various functionalities by adjusting the height of the flow channels, resulting in abrupt changes in height across different regions, which can cause flow losses. A more efficient approach would be to use continuously varying heights rather than abrupt transitions. This could potentially be achieved by employing conformal mapping techniques~\cite{conformal mapping-1,conformal mapping-2}.

In general, we propose a novel method for trapping moving particles within flowing fluids by creating a stationary region with no pressure gradient to capture the particles passing through it. Unlike traditional contactless tweezers that create a central potential well in a static environment to capture particles, our approach significantly expands the application scenarios of contactless tweezers. We achieve this by employing the principles of hydrodynamic metamaterials. Our tweezer design consists of two components: the core, which utilizes the zero-index material proposed in this study to immobilize particles, and the outer layer, which adjusts the flow field to ensure that the tweezer does not disturb the original flow. This capability is unique to our design and is unattainable by other contactless tweezers. The proposed tweezer operates passively by altering the height of the flow channel, eliminating the need for complex excitation devices. This passive operation was demonstrated experimentally, providing a clear and effective validation of the concept.

\medskip
\noindent\textbf{Acknowledgements}\par
\noindent We gratefully acknowledge funding from the National Natural Science Foundation of China (Grants Nos. 12035004, 12320101004, 12375040, 12088101, and U2330401) and the Innovation Program of Shanghai Municipal Education Commission (Grant No. 2023ZKZD06).

\medskip
\noindent\textbf{Compliance with ethics guidelines} \par 
\noindent The authors declare that they have no conflict of interest or financial conflicts to disclose.

\medskip
\noindent\textbf{Appendix A. Supplementary information} \par 
\noindent All data are available in the manuscript or the Supplementary information.

\end{document}